# Giant Intrinsic Carrier Mobilities in Graphene and Its Bilayer

S.V. Morozov[1,2], K.S. Novoselov[1], M.I. Katsnelson[3], F. Schedin[1], D.C. Elias[1], J.A. Jaszczak[4], A.K. Geim[1]

[1]Manchester Centre for Mesoscience and Nanotechnology,
University of Manchester, M13 9PL, Manchester, UK
[2]Institute for Microelectronics Technology, 142432 Chernogolovka, Russia
[3]Institute for Molecules and Materials, University of Nijmegen, 6525 ED Nijmegen, Netherlands
[4]Department of Physics, Michigan Technological University, Houghton, Michigan 49931, USA

*We have studied temperature dependences of electron transport in graphene and its bilayer and found extremely low electron-phonon scattering rates that set the fundamental limit on possible charge carrier mobilities at room temperature. Our measurements show that mobilities higher than 200,000 cm$^2$/Vs are achievable, if extrinsic disorder is eliminated. A sharp (threshold-like) increase in resistivity observed above ~200K is unexpected but can qualitatively be understood within a model of a rippled graphene sheet in which scattering occurs on intra-ripple flexural phonons.*

Graphene exhibits remarkably high electronic quality such that charge carriers in this one-atom-thick material can travel ballistically over submicron distances [1]. Electronic quality of materials is usually characterized by mobility $\mu$ of their charge carriers, and values of $\mu$ as high as 20,000 cm$^2$/Vs were reported for single-layer graphene (SLG) at low temperatures ($T$) [2-5]. It is also believed that $\mu$ in the existing samples is limited by scattering on charged impurities [6] or microscopic ripples [3,7]. Both sources of disorder can in principle be eliminated or reduced significantly. There are however intrinsic scatterers such as phonons that cannot be eliminated at room $T$ and, therefore, set a fundamental limit on electronic quality and possible performance of graphene-based devices. How high is the intrinsic mobility $\mu_{in}$ for graphene at 300K? This is one of the most important figures of merit for any electronic material but it has remained unknown.

In this Letter, we show that electron-phonon scattering in graphene and its bilayer is so weak that, if the extrinsic disorder is eliminated, room-$T$ mobilities ~200,000 cm$^2$/Vs are expected over a technologically relevant range of carrier concentration $n$. This value exceeds $\mu_{in}$ known for any other semiconductor [8]. In particular, our measurements show that away from the neutrality point (NP) resistivity $\rho$ of SLG has two components: in addition to the well-documented contribution $\rho_L=1/ne\mu$ due to long-range disorder [6,7], we have identified a small but notable $n$-independent resistivity $\rho_S$ indicating the presence of short-range scatterers [6,7,9]. We have also found that $\rho_L$ does not depend on $T$ below 300K, whereas $\rho_S$ exhibits a sharp rise above ~200K [10]. The latter contradicts to the existing theories [11] that expect a linear $T$ dependence. We attribute this behavior to flexural (out-of-plane) phonons [12] that are excited inside ripples. Bilayer graphene (BLG) samples exhibited no discernible $T$ dependence of $\mu$ away from NP, yielding even higher $\mu_{in}$. These findings provide an important benchmark for the research area and indicate that $\mu$ in graphene systems can be orders of magnitude higher than the values achieved so far. The reported measurements are also important for narrowing dominant scattering mechanisms in graphene, which remain hotly debated [3-7,11,13].

The studied devices were prepared from graphene obtained by micromechanical cleavage of graphite on top of an oxidized Si wafer (usually, 300nm of SiO$_2$) [14]. Single- and bi- layer crystallites were initially identified by their optical contrast [15], verified in some cases by Raman and atomic-force microscopy [2,14,16] and always crosschecked by measurements in high magnetic fields $B$, where SLG and BLG exhibited two distinct types of the quantum Hall effect [2,17]. To improve homogeneity, our standard Hall bar devices [1-3] were annealed at 200C in a H$_2$-Ar mixture [18] and, then, inside a measurement cryostat at 400 K in He. To avoid accidental breakdown, gate voltages $V_g$ were limited to ±50V ($n \propto \alpha V_g$ with $\alpha \approx 7.2 \times 10^{10}$cm$^{-2}$/V [1-5]). The measurements discussed below were carried out by the standard lock-in technique and refer to 7 SLG and 5 BLG devices with $\mu$ between 3,000 and 15,000 cm$^2$/Vs.

Figure 1 shows a characteristic behavior of $\rho(V_g)$ in SLG. The device exhibits a sharp peak close to zero $V_g$ (≈-0.2V), indicating little chemical doping [3]. Conductivity $\sigma =1/\rho$ is a notably sublinear function of $V_g$ in this device. Both linear and sublinear behaviors were reported previously [2-5]. To this end, if we subtract a constant resistivity $\rho_S$ (≈100 Ω in Fig. 1), then $\sigma_L = 1/[\rho(V_g) - \rho_S]$ becomes perfectly linear over the whole range of positive and negative $V_g$, except for the immediate vicinity of NP (<±3V). This linearization procedure was found to work extremely well for all our devices (the only exception was occasional devices with strongly distorted $\rho(V_g)$ indicating macroscopic inhomogeneity [1]). Furthermore, we digitized a number of curves in recent literature [4,5] and found the approach equally successful. This shows that resistivity of doped graphene can empirically be described by two contributions: $\rho_L \propto 1/n$ and $\rho_S$ independent of $n$ due to long- and short-range scatterers, respectively [3-7,9]. The latter contribution varies from sample to sample and becomes more apparent in high-$\mu$ samples. This observation resolves the controversy about the varying (linear vs sublinear) behavior reported in different experiments[2-5].

With this procedure in hand, it is now easier to describe $T$ dependence of graphene's conductivity. Figure 2 shows that $\sigma(V_g)$-curves become increasingly sublinear with increasing $T$. However, after linearization, the resulting curves (with $\rho_S$ subtracted) become essentially indistinguishable away from NP, collapsing onto a single curve $\sigma_L \propto |V_g|$ independently of $T$ (<300K; at higher $T$, we observed clear changes in the shape of $\sigma_L(V_g)$-curves, which indicates that the phonon contribution can no longer be described by $\rho_S$ independent on $n$). The extracted values of $\rho_S$ increase with $T$ as shown in Fig. 3 for 4 different devices. One can see that their $T$ dependent parts, $\Delta\rho_S =\rho_S(T)-\rho_S(0)$, behave qualitatively similar, despite different $\rho_S(0)$ at liquid-helium $T$. There is a slow (probably, linear) increase in $\rho_S$ at low $T$ but, above 200K, it rapidly shoots up (as $T^5$ or quicker). The latter $T$ dependence is inconsistent with scattering on acoustic phonons [11]. Note that $\Delta\rho_S$ does not exceed ≈50Ω at 300K, yielding $\mu_{in}$ between ~40,000 and 400,000 cm$^2$/Vs for characteristic $V_g$ between 5 and 50V ($n$ between ~3 and 30x10$^{11}$ cm$^{-2}$).

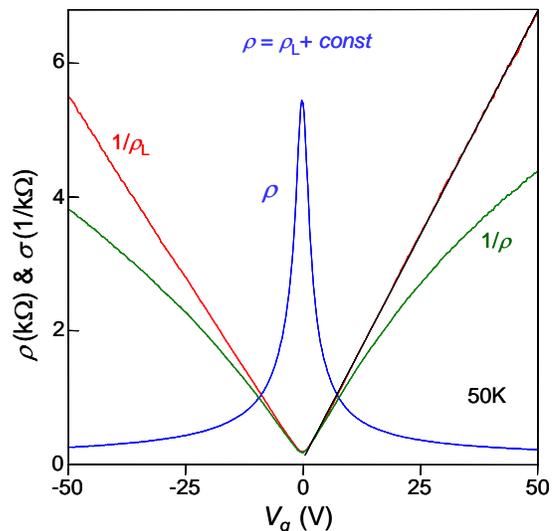

FIG 1. Resistivity $\rho$ (blue curve) and conducti-vity $\sigma=1/\rho$ (green) of SLG as a function of gate voltage. If we subtract a constant of ≈100 Ω (used here as a fitting parameter), the remaining part $\rho_L(V_g)$ of resistivity becomes inversely proportional to $V_g$ (red curve). The thin black line (on top of the red curve for $V_g$ >0) is to emphasize the linearity (the red curve is equally straight for negative $V_g$). The particular device was 1μm wide, and $T$ =50K was chosen to be high enough to suppress universal conductance fluctuations, still visible on the curves.

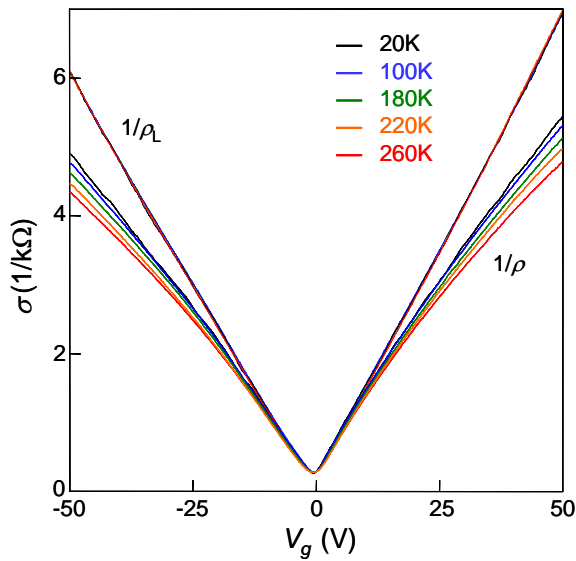

FIG 2. Electron transport in graphene below 300K can be described by the empirical expression $\rho(V_g,T)=\rho_L(V_g)+\rho_S(T)$ where $\rho_S$ is independent of $V_g$ but varies with $T$. After subtracting $\rho_S$ that for this sample changed from ≈40 Ω at low $T$ to ≈70 Ω at 260 K, the resulting curves $\sigma_L(V_g) = 1/\rho_L(V_g)$ became indistinguishable (the cluster marked $1/\rho_L$ consists of 5 such curves). The experiments were carried out in a field of 0.5T to ensure that weak localization corrections (rather small [1,22] but still noticeable) do not contribute into the reported $T$-dependences.

Now we turn to BLG. A typical behavior of its conductivity is shown in Fig. 4. BLG exhibits $\sigma(V_g)$ qualitatively similar to SLG's: away from NP, $\sigma_L \propto |V_g|$ yielding a constant $\mu$ of between 3,000 and 8,000 cm$^2$/Vs for our devices. This behavior (not reported before) is rather surprising because BLG's spectrum neither is similar to the conical spectrum of SLG [1,17] nor it can be considered parabolic (the measured cyclotron mass varies strongly with $n$ [19]). As for $T$ dependence in BLG, its only pronounced feature is a rapid decrease in $\rho$ around NP such that the peak value changes by a factor of 3 between liquid-helium to room $T$. The inset in Fig. 3 shows this dependence in more detail and compares it with the weak, non-monotonic behavior observed at NP in SLG. The origin of this pronounced difference between graphene and its bilayer lies in their different density of states near NP, which vanishes for SLG but is finite in BLG [1]. For SLG, the concentration of thermally excited carriers $\Delta n_T$ can be estimated as $(T/\hbar v_F)^2$ whereas in BLG it is $\sim Tm/\hbar^2$ ($v_F$ is the Fermi velocity in SLG and $m$ the effective mass in BLG [17,19]). In the latter case, $\Delta n_T \approx 10^{12}$ cm$^{-2}$ at room $T$, an order of magnitude larger than for SLG. The data in Fig. 3 (inset) are in agreement with this consideration. The weak $T$ dependence at NP is a unique feature of SLG and can be employed to distinguish SLG from thicker [14,20] crystallites.

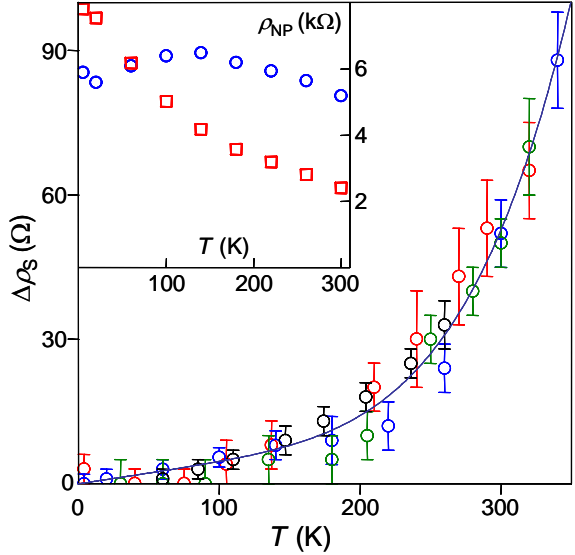

FIG 3. $T$-dependent part of resistivity for 4 SLG samples (symbols). The accuracy of measuring $\Delta\rho_S$ was limited by mesoscopic fluctuations at low $T$ and by gate hysteresis above 300K. The hysteresis appeared when $V_g$ was swept by more than 20V. To find $\Delta\rho_S$ at higher $T$, we recorded $\rho$ as a function of $n$ (found from simultaneous Hall measurements). The solid curve is the best fit by using a combination of $T$ and $T^5$ functions, which serves here as a guide to the eye. One of the samples (green circles) was made on top 200 nm of SiO$_2$ covered by further 100 nm of polymethylmethacrylate (PMMA) and exhibited $\mu \approx 7,000$ cm$^2$/Vs. The same values of $\mu$ for SiO$_2$ and PMMA substrates probably rule out charged impurities in SiO$_2$ as the dominant scattering mechanism for graphene. The inset shows $T$-dependence of maximum resistivity $\rho_{NP}$ (at the neutrality point) for SLG and BLG (circles and squares, respectively). Note a decrease in $\rho_{NP}$ with decreasing $T$ below 150 K for SLG, which is a generic feature seen in many samples.

Away from NP, we have never observed any sign of decrease in BLG's conductivity with increasing $T$. Comparison of Figs. 2 and 4 clearly illustrates that $T$-dependent scattering in BLG is substantially weaker than in SLG. Further measurements (Fig. 4) have shown that phonons do not contribute into BLG's $\mu$ within our experimental accuracy of <2%. This yields $\mu_{in} >300,000$ cm$^2$/Vs and a mean-free path of several microns at 300K.

Let us now try to understand the observed $T$ dependence of $\Delta\rho_S$ (Fig. 3). On one hand, it is partially consistent with scattering by in-plane phonons [11] in the sense that they lead to resistivity independent of $n$. On the other hand, such phonons give rise only to $\Delta\rho_S \propto T$, in clear disagreement with the measurements above 200K. These are rather general predictions and can be understood as follows. Because $v_F/v \approx 10^3$ ($v$ is the speed of sound), the Fermi wavelength $\lambda_F$ in our experiments exceeds the spatial scale associated with thermal phonons, $\approx 1/q_T$, at $T$ >10K ($q_T \approx T/v$ is the typical wavevector). This means that the scattering has a short-range character, leading to $\Delta\rho_S$ independent of $n$ [7,11]. Furthermore, the standard momentum and energy conservation considerations yield than only phonons with wavevectors $\sim k_F$ provide efficient (large-angle) scattering. The number of such phonons is $\propto T$ (given by the Boltzmann distribution in the limit $T \gg qv$) and, accordingly, $\Delta\rho_S \propto T$ [11] which can explain only our low-$T$ data. We also considered other $T$-dependent mechanisms such flexural phonons [12,21], electron-electron scattering and Umklapp processes, and they cannot explain the experimental behavior.

In the absence of a theory able to describe the rapid increase in $\Delta\rho_S$, we point out that the behavior is consistent with scattering on flexural phonons confined *within* ripples. Ripples are a common feature of cleaved graphene [18,22], suggesting that the atomic sheet is not fully bound to a substrate (as illustrated in ref. [23]) and, therefore, may exhibit local out-of-plane vibrations. First, because a characteristic size of ripples, $d \sim 10$ nm, is typically smaller than $\lambda_F$ [18,22], such vibrations induce predominantly short-range scattering. Second, at low $T$ ($q_T \ll 2\pi/d$), few flexural modes can be excited inside ripples but, as $T$ increases and typical wavelengths become shorter, more and more flexural phonons come into play. It was suggested [7] that electron scattering in graphene is dominated by static ripples quenched from the flexural-phonon disorder when graphene was deposited on a substrate at room $T$ (there are also short-range ripples induced by substrate's roughness [18]), which implies that any appreciable number of intra-ripple phonons start appearing only around room $T$. In fact, the observed behavior was predicted by Das Sarma and coworkers who – advocating for charged impurities as dominant scatterers in graphene [6,13] – noted that the model of quenched-ripple disorder [7] implied "strong temperature dependence (above a certain quenching temperature of about 100K) – an effect that has not been observed in the experiments" [13]. This is the experimental behavior reported here.

To estimate scattering rates $1/\tau$ for intra-ripple flexural phonons, one has to take into account two-phonon scattering processes because out-of-plane deformations modulate electron hopping only in the second order [7,12]. For our case of $q \gg k_F$, we have found [21]

$$\frac{1}{\tau} \cong \frac{\pi t'^2 k_F a^2}{8 v_F} \sum_{q \geq q_c} \frac{q^4}{M^2 \omega_q^2}\left(e^{\beta\omega_q} + \frac{2\beta\omega_q}{1 - e^{-2\beta\omega_q}}\right)\frac{1}{\left(e^{\beta\omega_q} - 1\right)^2}$$

where $\beta = \hbar/T$, $M$ is the mass of a carbon atom, $a$ the lattice constant, $\omega_q \propto q^2$ the flexural phonon frequency and $t'$ the derivative of the nearest-neighbor hopping integral with respect to deformation [7]. The integration goes over intra-ripple phonons that have $q$ larger than the cut-off wavevector $q_c \approx 2\pi/d$ imposed by the quenching. At low $T$, no flexural vibrations allowed inside ripples ($\delta\rho \approx 0$), whereas in the high-$T$ limit ($q_T \gg 2\pi/d$) the above expression allows the estimate $\delta\rho \approx (\hbar/e^2)(Td/2\pi\kappa a)^2$, yielding $\delta\rho \sim$ 100 to 1000 $\Omega$ at 300K ($\kappa \approx 1$ eV is the bending rigidity of graphene [7]). The rapid increase in $\Delta\rho_S$ above 200K can be attributed to transition between the low- and high-$T$ limits. The absence of any appreciable $T$ dependence in BLG is also consistent with the model, as BLG is more rigid and exhibits weaker rippling [22].

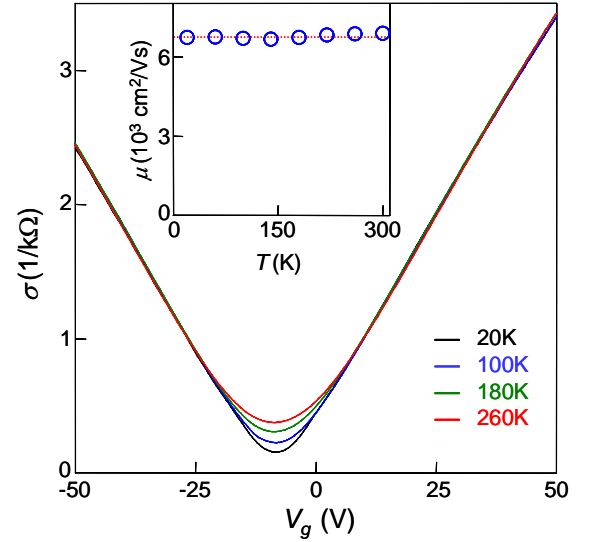

FIG 4. $T$ dependence in bilayer graphene. At the neutrality point, $\sigma$ rapidly increases with $T$ but, away from it, no changes are seen. $\sigma(V_g)$ exhibits a small sublinear contribution that can also be interpreted in terms of a constant $\rho_S$. The inset plots nominal values of $\mu$ found from linear fits of $\sigma_L(V_g)$ at $V_g >$ 20V away from NP and using $\rho_S \approx 50\Omega$. Here, $\mu$ does not change within $\approx 2\%$ and, if anything, shows a slight increase at higher $T$. The measurements were carried out at $B =$0.5T to suppress a small contribution of weak localization.

In summary, weak $T$ dependence of electron transport in graphene and its bilayer yields $\mu_{in} \sim$200,000 cm$^2$/Vs. The rapid rise in the small $T$-dependent part of $\rho$ in SLG lends support for the model of quenched-ripple disorder as an important scattering mechanism. The model suggests that the observed $T$ dependence is extrinsic and can probably be reduced together with ripples by depositing graphene on liquid-nitrogen-cooled substrates. If scattering on in-plane phonons does not increase in a flatter graphene sheet, $\mu_{in}$ could be truly colossal.

Acknowledgements: We are grateful to M. Dresselhaus who stimulated this work by repeatedly raising the question about graphene's intrinsic mobility. We also thank A. Castro Neto, S. Das Sarma, F. Guinea and V. Falko for useful discussions. This work was supported by EPSRC (UK) and the Royal Society.

*Note added at proof* – After the manuscript was submitted, Fratini and Guinea (arXiv:0711.1303v1) suggested that the observed strong $T$ dependence could alternatively be explained by scattering on surface phonons in the SiO$_2$ substrate, and this explanation was later used by Chen *et al* (arXiv:0711.3646) to analyze their experiment. The found agreement between the theory and both experiments is striking but let us note that two of the reported SLG samples were on top of 100 nm of PMMA (not SiO$_2$; see Fig. 3), and it would be fortuitous if the materials with so different polarizibility induce the same surface phonon scattering.